\documentclass[12pt]{article}
\usepackage[english]{babel}
\usepackage[latin1]{inputenc}
\usepackage{amsfonts,amssymb,amsmath, epsfig}
\usepackage{color,graphicx,graphics,psfrag}
\usepackage{amsmath,amstext,amssymb,amsfonts, amscd}
\usepackage{hyperref}          

\def\Box{\leavevmode\vbox{\hrule
     \hbox{\vrule\kern4pt\vbox{\kern4pt}%
           \vrule}\hrule}}
\def\blackbox{\leavevmode\vrule height 5pt width 4pt depth 0pt\relax}
\def\endproof{\null\hfill {$\blackbox$}\bigskip}

\newcounter{appendix}
\def\appendix{\advance\c@appendix by 1
   \def\thesection{\Alph{section}}
   \ifnum\c@appendix=1 \setcounter{section}{-1} \fi
   \@startsection {section}{1}{\z@}{-3.5ex plus -1ex minus 
   -.2ex}{2.3ex plus .2ex}{\Large\bf}}

\def\paragraph#1{{\bf #1\ }}

\newtheorem{lemma}{Lemma}[section]  

\newtheorem{theorem}[lemma]{Theorem}

\newtheorem{proposition}[lemma]{Proposition}

\newtheorem{remark}{Remark}[section]

\title{A Nash equilibrium macroscopic closure for kinetic models coupled with Mean-Field Games} 
\author{Pierre Degond $^{(1,2)}$, Jian-Guo Liu$^{(3)}$, Christian Ringhofer$^{(4)}$, } 
\date{} 
\begin{document}

\maketitle


\begin{center}
1-Université de Toulouse; UPS, INSA, UT1, UTM ;\\ 
Institut de Mathématiques de Toulouse ; \\
F-31062 Toulouse, France. \\
2-CNRS; Institut de Mathématiques de Toulouse UMR 5219 ;\\ 
F-31062 Toulouse, France.\\
email: pierre.degond@math.univ-toulouse.fr
\end{center}

\begin{center}
3- Department of Physics and Department of Mathematics\\
Duke University,
Durham, NC 27708, USA\\
email: jliu@phy.duke.edu
\end{center}

\begin{center}
4- School of Mathematics and Statistical Sciences, Arizona State University, Tempe AZ 85287, USA\\
email: ringhofer@asu.edu
\end{center}

\vspace{0.5 cm}
\begin{abstract}
We introduce a new mean field kinetic model for systems of rational agents
interacting in a game theoretical framework. This model is inspired from non-cooperative anonymous games with a continuum of players and Mean-Field Games. 
The large time behavior of the system is given by a macroscopic closure
with a Nash equilibrium serving as the local thermodynamic equilibrium. 
An application of the presented theory to a social model
(herding behavior) is discussed.
\end{abstract}

\medskip
\noindent
{\bf Acknowledgements:} This work has been supported by
KI-Net NSF RNMS grant No. 1107291. JGL and CR are greatful for the opportunity
to stay and work at the Institut de Mathématiques de Toulouse in fall 2012, under sponsoring of Centre National de la Recherche Scientifique and University Paul--Sabatier. The authors wish to thank A. Blanchet from University Toulouse 1 Capitole for enlighting discussions.


\vskip 0.4cm

\section{Introduction}
\label{sec:intro}

This paper is concerned with a dynamic model for an ensemble of rational agents or players in the game-theoretical sense. Each agent is endowed with two variables: a type variable $X$ which describes the state of the agent such as its position in some social or economic neighbourhood, its geographical position, etc., and a decision (or control or action or strategy) variable $Y$ which describes the degrees of freedom the agent can play with in the game. Each agents tries to minimize a cost function (or equivalently maximize a utility function) in the presence of the other players in the framework of a non-cooperative, anonymous game \cite{Green_Porter_Econometrica84, Schmeidler_JStatPhys73}. By contrast to equilibrium theory, we assume that the agents are not choosing the Nash equilibrium instantaneously \cite{Nash_PNAS50} (also known as Cournot-Nash equilibria), but rather work towards this goal by choosing the steepest descent direction towards the minimum of the cost function in each infinitesimal time step. In addition, this action is overlayed by some statistical noise, giving rise to Brownian fluctuations. This setup gives rise to a system of stochastic differential equations. 

We are interested in systems of a large number of agents where a continuum description can be adopted, in the way of games with a continuum of players \cite{Aumann_Econometrica64, Shapiro_Shapley_MathOperRes78, VonNeumann_Moregenstern_Princeton44} also known as Mean-Field Games \cite{Cardaliaguet_NotesCollegeFrance12, Lasry_Lions_JapanJMath07}. Indeed, in the situation of anonymous games with a large number of players, the construction of a mean field that serves as a mediator for describing inter-particle interactions constitutes an excellent approximation. In this kind of models, one describes the contribution of each particle to the creation of a mean field, and the effect of the mean field on each particle, by conceiving each particle as infinitesimal, i.e. by carrying out a kind of limit process on the number $N$ of particles for $N \rightarrow \infty$. We refer the reader to \cite{Blanchet_HDR12} for a nice introduction to game theory and Mean-Field Game. In the present paper we consider such mean field models, i.e. we consider a continuum of players which, under the usual molecular chaos assumptions, can be described by an effective equation for the probability distribution of single agents in the phase space $(x,y)$ consisting of the type and action variables.

In this large number of agents limit, a kinetic model for the time evolution of this probability distribution $f(x,y,t)$ can be written as follows. 
\begin {equation}\label {eq:intro_kinetic}
\partial_t f(x,y,t) + \nabla_x \cdot [ fV(x,y) ] -
\nabla_y \cdot [f\nabla _y\Phi _f(x,y,t)] = d \Delta_y f \ .
\end {equation}
The vector valued function $V(x,y)$ is given by the basic dynamics of the system, describing how the state $x$ evolves for a given control variable $y$. The forcing term $\nabla_y \cdot [f\nabla _y\Phi _f(x,y,t)]$ in equation (\ref {eq:intro_kinetic}) describes the agent trying to minimize the cost functional $\Phi _f$ by marching in the steepest descent direction $-\nabla _y\Phi _f$
(In the mean field model considered in this paper $\Phi _f$ will exhibit a functional dependence on the density $f$.) The Laplacian on the right hand side of equation (\ref {eq:intro_kinetic}) is a consequence of the Brownian noise in the system, with the diffusion coefficient $d$ corresponding to the variance. Again, by contrast to the usual Mean-Field Game models for agent systems \cite {Cardaliaguet_NotesCollegeFrance12, Lasry_Lions_JapanJMath07} where the optimum control is realized instantaneously (leading to the solution of a Hamilton - Jacobi - Bellman equation), our agents march an infinitesimal step towards the optimum at each time step by taking the steepest descent of the cost functional.

The substance of this paper is to relate the kinetic description given by equation (\ref {eq:intro_kinetic}) to the well-established game theoretical framework. We show that, in the homogeneous case, when the density $f$
is independent of the state $x$, steady state solutions of equation (\ref {eq:intro_kinetic}) correspond to Nash equilibria.
Therefore the model considered here consists of agents who try to achieve a Nash equilibrium by choosing their controls in
the direction of steepest descent towards this equilibrium. In the case of potential games \cite{Monderer_Shapley_GamesEconomicBehav96, Rosenthal_IntJGameTheory73}, when the cost functional $\Phi _f$ can be expressed as the functional derivative of a potential functional, we show that the kinetic equation can be expressed as the gradient flow of a free energy. Nash equilibria are the critical points of this free energy. Stable Nash equilibria are those which correspond to a global minimum of the free energy, the other minima corresponding to metastable ones. For special cases, we can prove that the dynamic solution of the kinetic equation converges to these stable or metastable Nash equilibrium solutions (see section \ref{subsec:social}). We note that potential games have originated from congestion games aimed at describing congested traffic situations \cite{Rosenthal_IntJGameTheory73} and that we will recover familiar models of traffic flow below (see Eq. \ref{eq:intro_hydro} and comments below). Recently, an approach similar to that developed here has been applied to pedestrian traffic \cite{Appert-Rolland_etal_preprint13_1, Appert-Rolland_etal_preprint13_2}. 

The main goal of the paper is to investigate the inhomogeneous case, when the probability density $f$ depends on the both the state (or type) $x$ and control variable $y$. We aim to derive macroscopic dynamic equations in the state variable $x$ only, which constitute good approximations to the solution of equation (\ref {eq:intro_kinetic}) at large scales. Indeed, we look at the system over time scales which are large compared to the typical time needed by the players to act on their control variables. Simultaneously, we suppose that the interactions between the players are localized in the state space $x$. This corresponds to a situation of so-called {\em bounded information} where the agents only take into account agents which are close to themselves in state space $x$ to make their decision, ignoring agents in the far field. In the macroscopic dynamics, we focus on scales in state space which are large compared to this interaction scale. Over these large time and state space scales, the distribution of agents in the control variable $y$ instantaneously realizes the local Nash equilibrium. This local Nash equilibrium describes the statistics of agents in control variable $y$ and depends on parameters which may vary over the large scale state variable $x$ and time $t$. Such parameters may be e.g. the local number density $\rho(x,t)$ of agents at given state $x$ and time $t$, or the mean or standard deviation of the local Nash equilibrium distribution.  

The resulting macroscopic equation represents, in the language of kinetic theory, the macroscopic closure of the kinetic equation, using the Nash equilibrium distribution as the Local Thermodynamic Equilibrium. For instance, the large time evolution of the density $\rho(x,t)$ is of the form
\begin {equation}\label {eq:intro_hydro}
\partial _t\rho (x,t)+\nabla _x\cdot (u\rho )=0\ ,
\end {equation}
where $\rho$ is related to the probability density $f$ by $\rho (x,t)=\int f(x,y,t)\ dy$. The macroscopic velocity $u(x,t)$ is given by the expectation of the local velocity $V(x,y)$ over the Nash equilibrium distribution. In the simplest possible case the macroscopic velocity $u$ can be expressed in terms of the macroscopic density $\rho$ as $u=u_\rho $, giving a closed (usually) hyperbolic conservation law such as in the Lighthill-Whitham-Richards model of traffic \cite{Lighthill_Whitham_ProcRoySocA55}. However, in many applications the structure of the macroscopic velocity $u$ is more complicated, and additional constitutive equations are needed to obtain a closed system from equation (\ref {eq:intro_hydro}). In such an occurrence a case by case study is necessary. We will give such an example in section \ref{subsec:social}.

This paper is organized as follows:
\begin {itemize}
\item[-]
In Section \ref {sec:kinetic} we define the basic model, consisting of a system
of stochastic differential equations, and state under what assumptions the solution
can be expressed in terms of a mean field density for one effective agent.
\item[-]
Section \ref {sec:homogeneous} is devoted to the analysis of equilibria.
For this purpose it is sufficient to consider the homogeneous case where the
density function $f$ in (\ref {eq:intro_kinetic}) does not depend on the state variable $x$.
In this case the equilibrium solution is given as the solution of a fixed point
problem and we show that this equilibrium solution is actually a Nash equilibrium
in the game theoretical sense. In the case of potential games, we provide a variational structure and Lypounov functional to  equation (\ref {eq:intro_kinetic}). 
\item[-]
Section \ref {sec:hydro} is concerned with the inhomogeneous case.
We consider the macroscopic limit in the regime
when the control variable $y$ is adjusted on a much faster time scale than
that of the evolution of the state variable $x$ and when the interactions are nearly local in state space. In the limit, this leads to
the macroscopic model (\ref {eq:intro_hydro}) where the macroscopic velocity $u$ has to be computed
from the local Nash equilibrium.
\item[-]
In Section \ref {subsec:social},  we apply the framework developed
so far to a model of social herding behavior,
where $V$ is an actual velocity in physical space, and the goal of each individual
is to adjust to the mean velocity of the ensemble.
Here, equilibrium distributions are given by the Von-Mises-Fischer distribution.
This serves as an example of a potential game.
However, the macroscopic limit equation (\ref {eq:intro_hydro}) is not well defined unless some additional
constitutive relations are used to determine the macroscopic velocity $u$. An other example pertaining to the evolution of the distribution of wealth in economic neighborhood can be found in \cite{Degond_etal_preprint13_1}. There are many models of social interactions and group formation based on a game theoretical approach (see e.g. \cite{Konishi_etal_GamesEconBehav97}). 
\item[-] In section \ref{sec:conclu}, a conclusion is given and perspectives are drawn. 
\end {itemize}

\setcounter{equation}{0}
\section{A mean-field model of social dynamics} 
\label{sec:kinetic}

We consider $N$ rational agents (or players) moving continuously in a space of social configurations ${\mathcal X}$. Each agent labeled $j$, $j \in \{ 1, \ldots, N\}$ has social configuration $X_j(t) \in {\mathcal X}$, depending on time $t \in {\mathbb R}_+$. It controls its state by an action (or decision) variable $Y_j(t)$ belonging to a space of control variables ${\mathcal Y}$. For simplicity, we suppose that ${\mathcal X} = {\mathbb R}^n$ and ${\mathcal Y}$ is a compact, orientable, connected manifold imbedded in ${\mathbb R}^p$ with or without boundary and endowed with the Riemannian structure induced by ${\mathbb R}^p$. Given $(X_j(t), Y_j(t))$, the $j$-th agent moves in configuration space with velocity $V_j(t) = V(X_j(t), Y_j(t))$, where $V=V(X,Y)$ is a given function of the configuration and decision variables. To act on their decision variables, the agents impose a given force $F_j^N$ which will be a function of his own and the other agents' configuration and decision variables. In addition to this force, each agent's decision variables are subject to Brownian noises which model uncertainties in the decision process as well as of the influence of the environment. In the game-theoretical literature, this is called a game with mixed strategy. Brownian noises for different agents are independent. In order to constrain the dynamics of the decision variables to the manifold ${\mathcal Y}$, the resulting combination of the force and Brownian noise is projected onto the tangent plane $T_{Y_j(t)}$ to ${\mathcal Y}$ at this point. The particle dynamics is given by the following Stochastic Differential Equation (SDE):
\begin{eqnarray}
&&\hspace{-1cm}
\dot X_j = V(X_j(t), Y_j(t)), \qquad dY_j = P_{T_{Y_j(t)}} \circ (F_j^N + \sqrt{2d} \, dW^j_t), 
\label{eq:discrete_dynamics}
\end{eqnarray}
where the dot indicates the time derivative, $P_{T_{Y_j(t)}}$ is the orthogonal projection onto the tangent plane to $Y_j(t)$, the symbol $\circ$ refers to the Stratonowich interpretation of the SDE, $dW^j_t$ for $j \in \{ 1, \ldots, N\}$ denote $N$ independent Brownian motions in ${\mathbb R}^p$ and $d$ is the diffusion coefficient. Finally $F_j^N$ denotes the force acting on the $j$-th agent which is described below. In the case where ${\mathcal Y}$ is a manifold with boundary, suitable boundary conditions must be given. Such boundary conditions will be specified later on in the kinetic framework (see Eq. (\ref{eq:bc})). That (\ref{eq:discrete_dynamics}) provides a well-defined SDE on ${\mathcal X} \times {\mathcal Y}$ follows from the theory descibed e.g. in \cite{Hsu_AMS02}.

We denote by $\vec{X}(t) =(X_1, \ldots, X_N)$, $\vec{Y}(t) =(Y_1, \ldots, Y_N)$ and $\hat Y_j = (Y_1, \ldots, Y_{j-1},$ $ Y_{j+1}, \ldots, Y_N)$. We also write ${\vec Y} = (Y_j, \hat Y_j)$ by abuse of notation. We assume the existence of a cost function $\Phi^N(\vec{X}, \vec{Y}, t)$, such that each agent tries to achieve a Nash equilibrium, i.e. relaxes its control variable $Y_j$ to an equilibrium $Y_j (\vec{X}, \hat Y_j,t)$ such that 
\begin{eqnarray}
&&\hspace{-1cm}
Y_j^N (\vec{X}, \hat Y_j,t) = \mbox{arg} \min_{Y_j \in {\mathcal Y}} \Phi^N(\vec{X}, Y_j, \hat Y_j, t), \quad \forall j \in \{ 1, \ldots, N\}. 
\label{eq:discrete_Nash}
\end{eqnarray}
Since such a goal cannot be achieved instantaneously, it chooses the steepest descent direction, i.e. it acts a force $F_j^N$ on itself such that 
\begin{eqnarray}
&&\hspace{-1cm}
F_j^N (\vec{X}, \vec{Y} ,t) = - \nabla_{Y_j} \Phi^N(\vec{X}, Y_j, \hat Y_j, t), \quad \forall j \in \{ 1, \ldots, N\}. 
\label{eq:steep_descent}
\end{eqnarray}
We now assume that $F_j^N$ is globally Lipschitz with respect to all its arguments, so that the system (\ref{eq:discrete_dynamics}) has global solutions. 

A more sophisticated way to optimize the action or control variables is to use a Hamilton-Jacobi-Bellman equation (see e.g. the Mean-Field Game theory of Lasry \& Lions \cite{Lasry_Lions_JapanJMath07}). We can also easily generalize this setting to a fiber bundle but we will stay in the frame of a trivial bundle (i.e. a cartesian produc) for simplicity.  

Now, we introduce the $N$-particle empirical distribution function 
$$ f^N(x,y,t) = \frac{1}{N} \sum_{j=1}^N \delta_{X_j(t)}(x) \otimes \delta_{Y_j(t)}(y), $$
and regard $f^N$ as a map from $t \in {\mathbb R}_+$ to $f^N(t) \in {\mathcal P}({\mathcal X} \times {\mathcal Y})$, where ${\mathcal P}({\mathcal X} \times {\mathcal Y})$ denotes the space of probability measures on ${\mathcal X} \times {\mathcal Y}$. 
We assume that in the mean-field limit $N \to \infty$ of the number of players going to infinity, there exists a one-particle distribution function $f = f(x,y,t)$, which maps $t \in {\mathbb R}_+$ to $f(t) \in {\mathcal P}_{\mbox{\scriptsize ac}}({\mathcal X} \times {\mathcal Y})$ where ${\mathcal P}_{\mbox{\scriptsize ac}}({\mathcal X} \times {\mathcal Y})$ is the space of probability measures on ${\mathcal X} \times {\mathcal Y}$ which are absolutely continuous with respect to the Lebesgue measure on ${\mathcal X} \times {\mathcal Y}$ (i.e. the measure on ${\mathcal X} \times {\mathcal Y}$ induced by the Lebesgue measure on ${\mathbb R}^n \times {\mathbb R}^p$), such that 
\begin{equation}
f^N \rightharpoonup f,
\label{eq:weak_cvg}
\end{equation}
in the weak star topology of bounded measures. We assume that a mean-field cost function exists. More precisely, we assume that there exists a map ${\mathcal P}_{\mbox{\scriptsize ac}}({\mathcal X} \times {\mathcal Y}) \to C^2({\mathcal X} \times {\mathcal Y})$, $f \mapsto \Phi_{f}$,   such that, for all trajectories $(X_j(t), Y_j(t))$ satisfying (\ref{eq:weak_cvg}), we have
\begin{eqnarray}
&&\hspace{-1.2cm}
\Phi^N(X_j(t), \hat X_j(t), Y_j(t), \hat Y_j(t), t) \to \Phi_{f(t)} (X_j(t), Y_j(t)) , \, \, \forall j \in \{ 1, \ldots, N\}, \, \, \forall t \geq 0.
\label{eq:mfl_pot}
\end{eqnarray}

Thanks to these assumption, in the limit $N \to \infty$, the one-particle distribution function $f$ is a solution of the following Fokker-Planck equation \cite{Sznitman_LN91}
\begin{eqnarray}
&&\hspace{-1cm}
\partial_t f + \nabla_x \cdot (V(x,y) f) + \nabla_y \cdot (F_f \, f) = d \Delta_y f, 
\label{eq:mfl_eq}
\end{eqnarray}
where $F_f = F_f(x,y,t)$ is given by 
\begin{eqnarray}
&&\hspace{-1cm}
F_f(x,y,t) = - \nabla_y \Phi_{f(t)} (x,y).
\label{eq:mfl_force}
\end{eqnarray}
In (\ref{eq:mfl_eq}), the symbol $\nabla_y \cdot$ denotes the divergence of tangent vector fields on ${\mathcal Y}$, while $\Delta_y$ is the Laplace-Beltrami operator on ${\mathcal Y}$. Below, we will also use $\nabla_y$ for the tangential gradient of functions defined on ${\mathcal Y}$. We supplement this system with an initial condition $f(0) = f_0$. For short, we will write $\Phi_{f(t)} = \Phi_f$. In the case where ${\mathcal Y}$ is a manifold with boundary, we set a zero flux condition on the boundary ${\mathcal X} \times \partial {\mathcal Y}$, namely: 
\begin{eqnarray}
&&\hspace{-1cm}
f \partial_n \Phi_f + d  \partial_n f = 0, \quad \mbox{ on } {\mathcal X} \times \partial {\mathcal Y}, 
\label{eq:bc}
\end{eqnarray}
where $\partial_n f(x,y)$ denotes the normal derivative of $f$ at $(x,y) \in {\mathcal X} \times \partial {\mathcal Y}$.

\setcounter{equation}{0}
\section{The homogeneous configuration case: convergence to Nash equilibria} 
\label{sec:homogeneous}

Here, we assume that the dynamics of the decision variables is independent of the state variables and we restrict the system to the decision variables only. In the kinetic-theory framework, this would refer to the spatially homogeneous case, where the spatial dependence is omitted. Then, $f$ becomes a mapping from $t \in [0,\infty[$ to $f(t) \in {\mathcal P}_{\mbox{\scriptsize ac}}({\mathcal Y})$, where ${\mathcal P}_{\mbox{\scriptsize ac}}({\mathcal Y})$ is now the space of absolutely continuous probability measures on ${\mathcal Y}$. The cost function $\Phi$ becomes a mapping from $f \in {\mathcal P}_{\mbox{\scriptsize ac}}({\mathcal Y})$ to $\Phi_f \in C^2({\mathcal Y})$. Eq. (\ref{eq:mfl_eq}) is now written:
\begin{eqnarray}
&&\hspace{-1cm}
\partial_t f = Q(f), \qquad Q(f) = \nabla_y \cdot \left(  f \, \nabla_y \Phi_f + d \nabla_y f \right) ,
\label{eq:mfl_eq_homo}
\end{eqnarray}
with initial condition given by $f_0$. We note that we can write the collision operator $Q(f)$ as follows:
\begin{eqnarray}
Q(f) &=& \nabla_y \cdot \Big( f \, \nabla_y \big( \Phi_f + d \ln f \big) \Big) = \nabla_y \cdot \big( f \, \nabla_y \mu_f \big) ,
\label{eq:Qwithmu}
\end{eqnarray}
with 
\begin{equation} 
\mu_f(y) = \Phi_f(y) + d \ln f(y). 
\label{eq:chempot_def}
\end{equation}
In the case where ${\mathcal Y}$ has boundary, then the boundary condition (\ref{eq:bc}) reduces to
\begin{eqnarray}
&&\hspace{-1cm}
f \, \partial_n \mu_f  = 0, \quad \mbox{ on } \partial {\mathcal Y},
\label{eq:bc_homo}
\end{eqnarray}
where $\partial_n \mu_f(y)$ is the normal derivative of $\mu_f$ at $y \in \partial {\mathcal Y}$. For a given function $\Phi(y)$, we introduce the Gibbs measure $M_\Phi(y)$ by:
\begin{eqnarray}
M_\Phi(y) &=& \frac{1}{Z_\Phi} \exp \big( - \frac{\Phi(y)}{d} \big), \qquad Z_\Phi = \int_{y \in {\mathcal Y}} \exp \big( - \frac{\Phi(y)}{d} \big) \, dy.
\label{eq:Von-Mises}
\end{eqnarray}
The definition of $Z_\Phi$ is such that $M_\Phi$ is a probability density, i.e. it satisfies 
\begin{eqnarray}
\int_{y \in {\mathcal Y}} M_\Phi(y) \, dy = 1. 
\label{eq:Von-Mises_norm}
\end{eqnarray}
Now, we can write
\begin{eqnarray}
Q(f) &=& d \,  \nabla_y \cdot \Big( M_{\Phi_f} \, \nabla_y \big(  \frac{f}{M_{\Phi_f}} \big) \Big) .
\label{eq:coll_operator}
\end{eqnarray}
We have the

\begin{lemma}
(i) For any sufficiently smooth function $f$ and $g$ on ${\mathcal Y}$, we have
\begin{eqnarray}
&&\hspace{-1cm}
\int_{y \in {\mathcal Y}} Q(f) \, \frac{g}{M_{\Phi_f}} \, dy = - d \, \int_{y \in {\mathcal Y}} \nabla_y \big( \frac{f}{M_{\Phi_f}} \big) \cdot \nabla_y \big( \frac{g}{M_{\Phi_f}} \big)  \, M_{\Phi_f} \, dy , 
\label{eq:dual}
\end{eqnarray}

\noindent
(ii) We have
\begin{eqnarray}
&&\hspace{-1cm}
\int_{y \in {\mathcal Y}} Q(f) \, \frac{f}{M_{\Phi_f}} \, dy = - d \, \int_{y \in {\mathcal Y}} \Big| \nabla_y \big( \frac{f}{M_{\Phi_f}} \big) \Big|^2  \, M_{\Phi_f} \, dy \leq 0.
\label{eq:dual_f=g}
\end{eqnarray}
\label{lem:dual}
\end{lemma}

\medskip
\noindent
{\bf Proof.} (i) Multiplying (\ref{eq:coll_operator}) by ${g}/{M_{\Phi_f}}$, integrating over $y$ and using Green's formula on ${\mathcal Y}$, we get (\ref{eq:dual}). In the application of Green's formula, the boundary term is either absent when ${\mathcal Y}$ has no boundary or vanishes due to the boundary condition (\ref{eq:bc_homo}) in the case where ${\mathcal Y}$ has a boundary. 
Indeed, we notice that 
\begin{eqnarray}
&&\hspace{-1cm}
\frac{f}{M_{\Phi_f}} = Z_{\Phi_f} \, e^{\frac{\mu_f}{d}}, \qquad  Q(f) = d \, Z_{\Phi_f} \,  \nabla_y \cdot \big( M_{\Phi_f} \, \nabla_y \,   e^{\frac{\mu_f}{d}} \big) .
\label{eq:f/M_Phi_f}
\end{eqnarray}
Therefore, the boundary term in Green's formula is written
$$ d \, Z_{\Phi_f} \, \int_{y \in \partial {\mathcal Y}} M_{\Phi_f}(y) \, \partial_n  \big(   e^{\frac{\mu_f(y)}{d}} \big)(y) \, \frac{g}{M_{\Phi_f}} \, dS(y) = 0, $$
where $dS(y)$ is the measure on $\partial {\mathcal Y}$ and where the integral is zero because 
$$\partial_n \big(  e^{\frac{\mu_f(y)}{d}} \big)(y) = \frac{1}{d} \, e^{\frac{\mu_f(y)}{d}} \, \partial_n \mu_f(y)  = 0, $$
by virtue of (\ref{eq:bc_homo}).  \\
(ii) We let $g=f$ in (\ref{eq:dual}) and get (\ref{eq:dual_f=g}). \endproof

\noindent
From Lemma \ref{lem:dual}, we deduce the following 

\begin{proposition}
The distribution function $f \in {\mathcal P}_{\mbox{\scriptsize ac}}({\mathcal Y})$ is an equilibrium solutions, i.e. a solution of $Q(f) = 0$ if and only if $f$ is of the form $f_{\mbox{\scriptsize{eq}}}$ where $f_{\mbox{\scriptsize{eq}}}$ is a solution of the following fixed point problem:
\begin{eqnarray}
&&\hspace{-1cm}
f_{\mbox{\scriptsize{eq}}} (y) = \frac{1}{Z_{\Phi_{f_{\mbox{\scriptsize{eq}}}}}} \exp \big( - \frac{\Phi_{f_{\mbox{\scriptsize{eq}}}}(y)}{d} \big), \qquad Z_{\Phi_{f_{\mbox{\scriptsize{eq}}}}} = \int_{y \in {\mathcal Y}} \exp \big( - \frac{\Phi_{f_{\mbox{\scriptsize{eq}}}}(y)}{d} \big) \, dy. 
\label{eq:mfl_equilibrium}
\end{eqnarray}
\label{prop:equilibrium}
\end{proposition}

\medskip
\noindent
{\bf Proof.} First, suppose that $Q(f) = 0$. Then, $\int_{y \in {\mathcal Y}} Q(f) \, \frac{f}{M_{\Phi_f}} \, dy = 0$. Therefore, thanks to (\ref{eq:dual_f=g}), $\frac{f}{M_{\Phi_f}}$ is a constant. Using the positivity of $M_{\Phi_f}$ and its normalization condition (\ref{eq:Von-Mises_norm}), we get $f = M_{\Phi_f}$. Consequently, for $f$ to be an equilibrium, it has to satisfy the fixed point problem (\ref{eq:mfl_equilibrium}). Conversely, if $f_{\mbox{\scriptsize{eq}}}$ is a solution of the fixed point problem (\ref{eq:mfl_equilibrium}), then $\frac{f_{\mbox{\scriptsize{eq}}}}{M_{\Phi_{f_{\mbox{\scriptsize{eq}}}}}} = 1$ and $Q(f_{\mbox{\scriptsize{eq}}}) = 0$ follows. \endproof

We now show that equilibria (\ref{eq:mfl_equilibrium}) are Nash equilibria for the mean-field game (also known as non-cooperative anonymous game with a continuum of players \cite{Cardaliaguet_NotesCollegeFrance12}) associated to the cost function $\mu_f(y)$. For such a game, a Nash equilibrium measure $f_{\mbox{\scriptsize{NE}}} \in {\mathcal P}({\mathcal Y})$ is such that \cite{Cardaliaguet_NotesCollegeFrance12} (see also \cite{Blanchet_Carlier, Blanchet_Mossay_Santambrogio}) there exists a constant $K$ and
\begin{equation} \left\{ \begin{array}{ll} \mu_{f_{\mbox{\tiny{NE}}}}(y) = K & \qquad \forall y \in \mbox{Supp} (f_{\mbox{\tiny{NE}}}), \\
\mu_{f_{\mbox{\tiny{NE}}}}(y) \geq K & \qquad \forall y \in {\mathcal Y}. \end{array} \right. 
\label{eq:NE}
\end{equation}
This definition is equivalent to the following statement \cite{Cardaliaguet_NotesCollegeFrance12}:
\begin{equation} 
\int_{y \in {\mathcal Y}} \mu_{f_{\mbox{\tiny{NE}}}}(y) \, f_{\mbox{\tiny{NE}}}(y) \, dy = \inf_{f \in {\mathcal P}_{\mbox{\scriptsize{ac}}}({\mathcal Y})} \int_{y \in {\mathcal Y}} \mu_{f_{\mbox{\tiny{NE}}}}(y) \, f(y) \, dy .
\label{eq:NE2}
\end{equation}
Eq. (\ref{eq:NE2}) is called the 'mean-field' equation. Now, we have the following

\begin{theorem}
Let $f \in {\mathcal P}_{\mbox{\scriptsize ac}}({\mathcal Y})$. Then the two following statements are equivalent:\\
\indent (i) $f$ is an equilibrium (\ref{eq:mfl_equilibrium}), \\
\indent (ii) $f$ is a Nash equilibrium (\ref{eq:NE}).
\label{thm:NE}
\end{theorem}

\noindent
{\bf Proof.} (i) $\Rightarrow$ (ii). Let $f_{\mbox{\scriptsize{eq}}}$ be an equilibrium (\ref{eq:mfl_equilibrium}). Since ${\mathcal Y}$ is compact and $\Phi_f$ is continuous on ${\mathcal Y}$ for any $f \in {\mathcal P}_{\mbox{\scriptsize ac}}({\mathcal Y})$, then $\Phi_{f_{\mbox{\scriptsize{eq}}}}$ is bounded. Therefore, its support is the entire manifold ${\mathcal Y}$ and the second line of (\ref{eq:NE}) reduces to the first line. We easily compute that $K = - d \ln Z_{\Phi_{f_{\mbox{\scriptsize{eq}}}}}$. Therefore, $f_{\mbox{\scriptsize{eq}}}$ is a Nash equilibrium (\ref{eq:NE}). 

\medskip
\noindent
(ii) $\Rightarrow$ (i). Let $f_{\mbox{\tiny{NE}}}$ be a Nash equilibrium (\ref{eq:NE}). We show that $\mbox{Supp} (f_{\mbox{\tiny{NE}}}) = {\mathcal Y}$. Indeed, by contradiction, suppose $\mbox{Supp} (f_{\mbox{\tiny{NE}}}) \varsubsetneq {\mathcal Y}$. There exists $y \in {\mathcal Y}$ such that $f_{\mbox{\tiny{NE}}}(y) = 0$. Then, because of the log inside (\ref{eq:chempot_def}) and the boundedness of $\Phi_{f_{\mbox{\tiny{NE}}}}$, we have $\mu_{f_{\mbox{\tiny{NE}}}} (y) = - \infty$ which is a contradiction to the second line of (\ref{eq:NE}). Therefore, by the first line of (\ref{eq:NE}),  $\mu_{f_{\mbox{\tiny{NE}}}}$ is identically constant over the entire space ${\mathcal Y}$. From the expression of $\mu_{f_{\mbox{\tiny{NE}}}}$ in (\ref{eq:chempot_def}), $ f_{\mbox{\tiny{NE}}}$ is proportional to $\exp (- {\Phi_{f_{\mbox{\tiny{NE}}}}}/{d} )$, which means that it is an equilibrium (\ref{eq:mfl_equilibrium}). 
\endproof

The mean-field model (\ref{eq:mfl_eq_homo}), (\ref{eq:Qwithmu}) can be recast as a transport equation as follows
\begin{eqnarray}
&&\hspace{-1cm}
\partial_t f + \nabla_y \cdot (v \, f) = 0, 
\label{eq:mfl_transport} \\
&&\hspace{-1cm}
v = - \nabla_y \mu_f. 
\label{eq:mfl_drift} 
\end{eqnarray}
It describes the bulk motion of agents which move in the direction of the steepest descent towards the minimum of $\mu_f$. When all agents have reached the minimum of $\mu_f$, then $\mu_f$ is a constant and describes a Nash Equilibrium. Therefore, in the proposed dynamics, the agents choose as their action to move in the steepest descent direction towards the Nash equilibrium. 

\begin{remark}
{\bf Variational structure and potential games \cite{Monderer_Shapley_GamesEconomicBehav96}.} We suppose that there exists a functional ${\mathcal U} (f)$ such that 
$$ \Phi_f(y) = \frac{\delta {\mathcal U} (f)}{\delta f}(y), \quad \forall y \in {\mathcal Y}, $$
where $\frac{\delta {\mathcal U} (f)}{\delta f}$ is the functional derivative of ${\mathcal U}$ defined by 
\begin{equation}
\int_{y \in {\mathcal Y}} \frac{\delta {\mathcal U} (f)}{\delta f}(y) \, \phi(y) \, dy = \lim_{s \to 0} \frac{1}{s} ({\mathcal U} (f + s \phi) - {\mathcal U} (f)), 
\label{eq:fct_der}
\end{equation}
for any test function $\phi(y)$. We note that the existence of such a functional gives a very strong constraint on $\Phi$ (if $f$ were finite-dimensional, i.e. if ${\mathcal Y}$ were replaced by a finite set and eq. (\ref{eq:mfl_eq_homo}) by a system of ordinary differential equations, that would mean that $\Phi$ is the gradient of the scalar potential ${\mathcal U}$). We will call ${\mathcal U}$ the potential energy. A game associated to such a cost function $\Phi$ is called a potential game \cite{Monderer_Shapley_GamesEconomicBehav96}. We now introduce the entropy functional:
\begin{equation} 
{\mathcal S} (f) = d \, \int_{y \in {\mathcal Y}} f(y) \, \ln f(y) \, dy, 
\label{eq:entropy_def}
\end{equation}
and the free energy 
\begin{equation} 
{\mathcal F} (f) = {\mathcal U} (f)  + {\mathcal S} (f).   
\label{eq:freeener_def}
\end{equation}
It is a simple matter to find that 
$$ \frac{\delta {\mathcal S} (f)}{\delta f}(y)  = d  \ln f(y) .  $$
Therefore, we have 
\begin{equation}
\frac{\delta {\mathcal F} (f)}{\delta f}(y)  = \Phi_f(y)  + d  \ln f(y) = \mu_f(y)  ,  
\label{eq:def_mu}
\end{equation}
so that, in this setting, $\mu_f(y)$ can be seen as the 'chemical potential' associated to the free energy ${\mathcal F} (f)$. Now, we can recast (\ref{eq:mfl_eq_homo}), (\ref{eq:Qwithmu}) as 
\begin{eqnarray}
&&\hspace{-1cm}
\partial_t f = \nabla_y \cdot \big(  (\nabla_y \mu) f \big) = \nabla_y \cdot \Big(  \nabla_y \big( \frac{\delta {\mathcal F} (f)}{\delta f} \big) \, f \Big).
\label{eq:mfl_eq_homo_var}
\end{eqnarray}
But, for a function $f(y,t)$, we have
$$ \frac{d}{d t} {\mathcal F} (f(\cdot, t)) = \int_{y \in {\mathcal Y}} \frac{\delta {\mathcal F} (f(\cdot,t))}{\delta f}(y) \, \frac{\partial f}{\partial t} (y,t) \, dy . $$
Inserting (\ref{eq:mfl_eq_homo_var}) into the above equation, and using Green's formula, we have:
$$ \frac{d}{d t} {\mathcal F} (f(\cdot, t)) = - \int_{y \in {\mathcal Y}} f(y,t) \, \Big| \nabla_y \frac{\delta {\mathcal F} (f(\cdot,t))}{\delta f}(y) \Big|^2 (y,t) \, dy : = - {\mathcal D}(f(\cdot,t)) \leq 0. $$
Therefore, ${\mathcal F}$ is Liapounov functional for this dynamic and ${\mathcal D}$ is the free-energy dissipation term. By fine analysis of ${\mathcal D}$, it is possible in some cases to deduce decay rates from this kind of estimate \cite{Frouvelle_Liu_SIMA12, Villani_AMS03}. Equilibria given by (\ref{eq:mfl_equilibrium}) are critical points of ${\mathcal F}$ subject to the constraint $\int f \, dy = 1$ and the chemical potential $\mu$ is the Lagrange multiplier of this constraint in this optimization problem. Each of these critical points corresponds to a Nash equilibrium. However, these critical points are not necessarily global minimizers of the free energy. Among these equilibria, the ground states, which are the global minimizers of ${\mathcal F}$ are the most stable ones. Other equilibria are either not stable or only locally stable (or meta-stable). The co-existence of several stable equilibria may give rise to phase transitions and hysteresis behavior if bifurcation parameters are involved and varied. 
\label{rem:variational}
\end{remark}

\setcounter{equation}{0}
\section{The inhomogeneous configuration case:  Nash Equilibrium macroscopic closure}
\label{sec:hydro}

Now, we return to the inhomogeneous configuration case (\ref{eq:mfl_eq}), (\ref{eq:mfl_force}) where the positions of the players in the social configuration space is considered. The goal of this section is to investigate the ensemble motion of the players at large time scales, averaging out over their individual decision variables. For this purpose, we have to assume a temporal scale separation, where individual decisions are fast compared to the evolution of the ensemble of players in configuration space. We also need to observe the system as a bulk, averaging out the fine details of the individual players in configuration space. Therefore, we will introduce a suitable coarse-graining procedure. Taking advantage that at large times, individuals relax their decision variables towards that corresponding to a global Nash equilibrium given by (\ref{eq:mfl_equilibrium}), we use this equilibrium as a prescription for the internal decision variable distribution of the agents. In this section, we provide the details of this coarse-graining process, known as the hydrodynamic limit in kinetic theory.  

In order to manage the various scales in a proper way, we first change the variables to dimensionless ones. Let $t_0$ be a time unit and let $x_0 = a t_0$, where $a$ is the typical magnitude of $V$. We choose $t_0$ in such a way that the magnitude of  $\Phi$ is ${\mathcal O}(1)$ and introduce the quantity $\tilde d = d t_0 = {\mathcal O}(1)$. The decision space ${\mathcal Y}$ is already dimensionless and the variable $y$ does not require any scaling. Introducing new variables $\tilde x = x/x_0$, $\tilde t = t/t_0$, $\tilde f (\tilde x, y, \tilde t)= x_0^n f(x,y,t)$, $\tilde V(\tilde x, y) = V(x,y) / a$, $\tilde \Phi_{\tilde f}(\tilde x,y) = \Phi_f(x,y)$, Eq. (\ref{eq:mfl_eq}) is written:
\begin{eqnarray}
&&\hspace{-1cm}
\partial_{\tilde t} {\tilde f} + \nabla_{\tilde x} \cdot (\tilde V(\tilde x,y) {\tilde f}) + \nabla_y \cdot ({\tilde F}_{\tilde f} \, {\tilde f}) = \tilde d \Delta_y {\tilde f}, 
\label{eq:mfl_eq_tilde}
\end{eqnarray}
where ${\tilde F}_{\tilde f}$ is given by 
\begin{eqnarray}
&&\hspace{-1cm}
{\tilde F}_{\tilde f}({\tilde x},y,{\tilde t}) = - \nabla_y {\tilde \Phi}_{\tilde f} (\tilde x,y), 
\label{eq:mfl_force_tilde}
\end{eqnarray}

We now introduce the macroscopic scale. Refering to the discussion of the beginning of this section, we change the configuration space unit and the time unit to new ones $x'_0$, $t'_0$ which are large compared to $x_0$, $t_0$. Specifically, we let $\varepsilon \ll 1$ be a small parameter and define $x'_0 = x_0 / \varepsilon$, $t'_0 = t_0 / \varepsilon$. Here, $\varepsilon$ refers to the "small" average change of the configuration of the ensemble of agents on the "fast" time-scale of the evolution of the decision variables. In the kinetic framework, $\varepsilon$ would be a measure of the particle mean-free path in macroscopic units. By doing so, we change the space and time variables $\tilde x$ and $\tilde t$ to macroscopic variables $\hat x = \varepsilon \tilde x$, $\hat t = \varepsilon \tilde t$ and define $\hat f (\hat x, y, \hat t) = \varepsilon^{-n} \tilde f (\tilde x, y, \tilde t)$. Inserting this change of variables into (\ref{eq:mfl_eq_tilde}), (\ref{eq:mfl_force_tilde}), we are led to the following perturbation problem (dropping the hats and tildes for simplicity):
\begin{eqnarray}
&&\hspace{-1cm}
\varepsilon \big( \partial_t f^\varepsilon + \nabla_x \cdot (V(x,y) f^\varepsilon) \big) + \nabla_y \cdot (F^\varepsilon_{f^\varepsilon} \, f^\varepsilon) = d \Delta_y f^\varepsilon, 
\label{eq:mfl_eq_eps}
\end{eqnarray}
where $F_{f^\varepsilon}$ is given by 
\begin{eqnarray}
&&\hspace{-1cm}
F^\varepsilon_f(x,y,t) = - \nabla_y \Phi^\varepsilon_f(x,y). 
\label{eq:mfl_force_eps}
\end{eqnarray}

For a distribution function $f(x,y)$, $f \in {\mathcal P}_{\mbox{\scriptsize ac}}({\mathcal X} \times {\mathcal Y})$, we assume that $\Phi^\varepsilon_f$ can be developed as follows:
\begin{eqnarray}
&&\hspace{-1cm}
\Phi^\varepsilon_{f}(x,y) = \Phi_{\rho(x),\, \nu_x} (x,y) + {\mathcal O}(\varepsilon^2),
\label{eq:Phi_eps_exp}
\end{eqnarray}
with 
\begin{eqnarray}
&&\hspace{-1cm}
\nu_x(y) = \frac{1}{\rho(x)} f(x,y), \qquad \rho(x) = \int_{y \in {\mathcal Y}} f(x,y) \, dy, 
\label{eq:mfl_mom_eps}
\end{eqnarray}
and $\Phi_{\rho, \nu}$ is a map $[0,\infty) \times {\mathcal P}_{\mbox{\scriptsize ac}}({\mathcal Y}) \to C^2({\mathcal X} \times {\mathcal Y})$, $(\rho, \nu) \mapsto \Phi_{\rho, \, \nu}$. In short, $\nu_x$ is the conditional probability density of $f$ conditionned by fixing the position $x \in {\mathcal X}$ and it belongs to ${\mathcal P}_{\mbox{\scriptsize ac}}({\mathcal Y})$. Eq. (\ref{eq:Phi_eps_exp}) states that, up to factors of order ${\mathcal O}(\varepsilon^2)$, the cost function is a functional of this conditional probability and of the density only, and therefore, only depends on local quantities at social position $x$. The  ${\mathcal O}(\varepsilon^2)$ term collects all non-local effects in social position space. These effects are supposed to be much smaller than the local ones. This is an expression of the scale separation in social space: local effects in social space are supposed to have a much bigger influence that non-local ones on a given subject.

The macroscopic limit is about taking the limit $\varepsilon \to 0$ in this set of equations. 
In order to do so, we write (\ref{eq:mfl_eq_eps}), (\ref{eq:mfl_force_eps}) as follows:
\begin{eqnarray}
&&\hspace{-1cm}
\partial_t f^\varepsilon + \nabla_x \cdot (V(x,y) f^\varepsilon) = \frac{1}{\varepsilon} Q(f^\varepsilon),
\label{eq:mfl_eq_eps_2}
\end{eqnarray}
with $Q$ given by 
\begin{eqnarray}
&&\hspace{-1cm}
Q(f) = \nabla_y \cdot \left(  \nabla_y \Phi_{\rho(x,t), \, \nu_{x, t}} \, f + d \nabla_y f \right) ,
\label{eq:Q_def}
\end{eqnarray}
and where $\rho(x,t)$, $\nu_{x, t}$ are related to $f(t)$ by (\ref{eq:mfl_mom_eps}). Here, we have used (\ref{eq:Phi_eps_exp}) to replace $\Phi^\varepsilon$ by $\Phi$ in (\ref{eq:mfl_eq_eps_2}), (\ref{eq:Q_def}) and dropped the remaining ${\mathcal O}(\varepsilon)$ terms. In the example section below, we will show that this assumption is actually quite natural.

Here again, we emphasize that $(x,t)$ now refers to slow variables. The left-hand side of (\ref{eq:mfl_eq_eps_2}) describes how the distribution of agents as a function of the external variables (the social configuration $x$) evolves. This evolution is driven by the the fast, local evolution of this distribution as a function of the individual decision variables $y$ described by the right-hand side. The parameter $\varepsilon$ at the denominator highlights that fact that the internal decision variables evolve on a faster time-scale than the external social configuration variables. The fast evolution of the internal decision variables drives the system towards an equilibrium, i.e.,  solution of $Q(f) = 0$. Such a solution is referred to in physics as a Local Thermodynamical Equilibrium (LTE). Below, we use the results of the previous section to show that, in this case, the LTE's are given by Nash equilibria. 

To highlight this fact, by factoring out $\rho(x,t)$ from the expression of $Q$ in (\ref{eq:Q_def}), we can recast it as follows: 
\begin{eqnarray}
&&\hspace{-1cm}
Q(f) := \rho(x,t) \, {\mathcal Q}_{\rho(x,t)} (\nu_{x,t}) , 
\label{eq:Q_def_nu}
\end{eqnarray}
where, for any $\rho \in {\mathbb R}^+$, we define the operator ${\mathcal Q}_{\rho}$ acting on ${\mathcal P}_{\mbox{\scriptsize ac}}({\mathcal Y})$ as follows: 
\begin{eqnarray}
&&\hspace{-1cm}
{\mathcal Q}_{\rho}(\nu) := \nabla_y \cdot \left(  \nabla_y \Phi_{\rho, \, \nu} \, \nu + d \nabla_y \nu \right) .
\label{eq:Q_def_rho}
\end{eqnarray}
The equation $Q(f)=0$ can then be recast (supposing that $\rho(x,t) \not = 0$) into ${\mathcal Q}_{\rho(x,t)}(\nu_{x,t}) = 0$. But the operator ${\mathcal Q}_{\rho}$ freezes the slow variables $(x,t)$ and acts only on the distribution of agents in the decision variable $y$. Therefore, this equation is merely a homogeneous configuration problem and we can apply Proposition \ref{prop:equilibrium} to solve it. This leads to the following lemma whose proof is a direct application of Proposition \ref{prop:equilibrium} and is omitted. 

\begin{lemma}
The LTE, i.e. the solutions of $Q(f)= 0$ are given by 
\begin{eqnarray}
&&\hspace{-1cm}
f(x,y,t) = \rho(x,t) \, \nu_{\mbox{\scriptsize{eq}}, \, \rho(x,t)} (y),
\label{eq:equi_nonhomo}
\end{eqnarray}
where $\nu_{\mbox{\scriptsize{eq}}, \, \rho} (y)$ is a solution of ${\mathcal Q}_{\rho} (\nu) = 0$. Such solutions $\nu_{\mbox{\scriptsize{eq}}, \, \rho} (y)$ are given by the resolution of the Nash equilibrium fixed point problem
\begin{eqnarray}
&&\hspace{-1.2cm}
\nu_{\mbox{\scriptsize{eq}}, \, \rho} (y) = \frac{1}{Z_{\Phi_{\rho, \, \nu_{\mbox{\tiny{eq}}, \rho}}}} \exp \big( - \frac{\Phi_{\rho, \, \nu_{\mbox{\tiny{eq}}, \rho}}(y)}{d} \big), \quad Z_{\Phi_{\rho, \, \nu_{\mbox{\tiny{eq}}, \rho}}} = \int_{y \in {\mathcal Y}} \exp \big( - \frac{\Phi_{\rho, \, \nu_{\mbox{\tiny{eq}}, \rho}}(y)}{d} \big) \, dy. 
\label{eq:mfl_equilibrium_inhomo}
\end{eqnarray}
\label{lem:LTE_inhomo}
\end{lemma}

Now, we can state the result for the coarse-graining limit $\varepsilon \to 0$ inside Eq. (\ref{eq:mfl_eq_eps_2}). We have

\begin{theorem}
Suppose that the solution $f^\varepsilon$ to (\ref{eq:mfl_eq_eps_2}) converges to a function $f$ when $\varepsilon \to 0$ smoothly, which means in particular that all derivatives of $f^\varepsilon$ converge to the corresponding derivative of $f$. Then, formally $f$ is given by an LTE (\ref{eq:equi_nonhomo}).
The density $\rho(x,t)$ satisfies the following conservation law:
\begin{eqnarray}
&& \hspace{-1cm} \partial_t \rho + \nabla_x \cdot (\rho u) = 0 , \label{eq:mass} 
\end{eqnarray}
with 
\begin{equation}
u = u[x,\nu_{\mbox{\scriptsize{eq}}, \, \rho(x,t)}], 
\label{eq:mean_vel_equi}
\end{equation}
being the mean velocity of $\nu_{\mbox{\scriptsize{eq}}, \, \rho(x,t)}$ and $u[x,\nu]$ is given by
\begin{eqnarray}
&& \hspace{-1cm} u[x,\nu] = \int_{y \in {\mathcal Y}} V(x,y) \, \nu(y) \, dy,
\label{eq:u} 
\end{eqnarray}
for all distributions $\nu \in {\mathcal P}_{\mbox{\scriptsize ac}}({\mathcal Y})$.
\label{thm:inhomogeneous}
\end{theorem}

\medskip
\noindent
{\bf Proof.} From (\ref{eq:mfl_eq_eps_2}), we have that $Q(f^\varepsilon) = {\mathcal O}(\varepsilon)$ and owing to the convergence assumptions made on $f^\varepsilon$, we have $Q(f) = 0$. Thanks to Lemma \ref{lem:LTE_inhomo}, $f$ is of the form (\ref{eq:equi_nonhomo}). Now, observe that $1$ is a collisional invariant of $Q$, meaning that 
$ \int_{y \in {\mathcal Y}} Q(f) (y) \, dy = 0, $
for all functions $f(y)$ (simply by Green's formula and the boundary conditions). Therefore, integrating (\ref{eq:mfl_eq_eps_2}) with respect to $y$ leads to
\begin{eqnarray}
&& \hspace{-1cm} \partial_t \rho^\varepsilon + \nabla_x \cdot (\rho^\varepsilon u^\varepsilon) = 0 , \label{eq:mass_eps} 
\end{eqnarray}
with $u^\varepsilon(x,t) = u[x,\nu^\varepsilon_{x,t}]$ where $u[x,\nu]$ is given by (\ref{eq:u}) and $\nu^\varepsilon_{x,t}$ is related to $f^\varepsilon$ by the first eq. (\ref{eq:mfl_mom_eps}).  
Then, taking the limit $\varepsilon \to 0$ in (\ref{eq:mass_eps}) and using that $\rho^\varepsilon \to \rho$ and $ \nu^\varepsilon_{x,t} \to \nu_{\mbox{\scriptsize{eq}}, \, \rho(x,t)}$, we get that $u^\varepsilon \to u[x,\nu_{\mbox{\scriptsize{eq}}, \, \rho(x,t)}]$ and that the limit of Eq. (\ref{eq:mass_eps}) is precisely (\ref{eq:mass}). \endproof

\begin{remark}
We note that Eq. (\ref{eq:mass}) (complemented with an initial condition $\rho_0(x)$ and possibly boundary conditions) does not necessarily lead to a closed system. We will provide examples in the next section where additional equations may be required to provide a closed problem. However, in many cases, the solution of the Nash equilibrium problem (\ref{eq:mfl_equilibrium}) is not known explicitly. This suggests the development of coarse-graining strategies, based on e.g. the Heterogeneous Multiscale Method \cite{E_Cambridge11} or kinetic upscaling \cite{Degond_etal_MMS06}. 
\label{rem:not_closed}
\end{remark}

\setcounter{equation}{0}
\section{Models of social herding behavior}
\label{subsec:social}

\subsection{General framework}
\label{subsec:social_framework}

Here, we specify the potential $\Phi_f(x,y)$ as given by the following kernel:
\begin{eqnarray}
&& \hspace{-1cm} \Phi_f(x,y) = \int_{(x',y') \in {\mathcal X}\times {\mathcal Y}} k(x,y,x',y') f(x',y') \, dx' \, dy' , 
\label{eq:kernel} 
\end{eqnarray}
where $(x,y,x',y') \in ({\mathcal X} \times {\mathcal Y})^2  \mapsto k(x,y,x',y') \in{\mathbb R}$ is a given function. 
To be more specific, we focus on a model of social herding behavior, where pairs of agents try to minimize the angle between their respective social velocities. Namely, we set: 
\begin{eqnarray}
&& \hspace{-1cm} k(x,y,x',y')  = - K(x,x') \, \, V(x,y) \cdot V(x',y'), 
\label{eq:kernel_separate} 
\end{eqnarray}
where $V(x,y)$ is the velocity in social space specified earlier and the dot refers to inner product in the vector space ${\mathcal X}$. By trying to minimize the angle between their own velocity and that of their neighbors, the agents adopt a mimetic behavior, and tend to move in social space in the same direction as the others. We can write 
\begin{eqnarray}
&& \hspace{-1cm} \Phi_f(x,y) = - V(x,y) \cdot {\mathcal W}_f(x), 
\label{eq:kernel_2} 
\end{eqnarray}
with
\begin{eqnarray}
&& \hspace{-1cm} {\mathcal W}_f(x) = \int_{(x',y') \in {\mathcal X}\times {\mathcal Y}} K(x,x') \, V(x',y') \,  f(x',y') \, dx' \, dy'  \in {\mathcal X}. 
\label{eq:kernel_3} 
\end{eqnarray}
We can view ${\mathcal W}_f$ as some average of  $V(x,y)$ over $f$. 

Now, let us first focus on the homogeneous configuration case, letting $K=1$. In this case, $f(y,t)$ satisfies (\ref{eq:mfl_eq_homo}) with $\Phi_f(y)$ given by:
\begin{eqnarray}
&& \hspace{-1cm} \Phi_f(y) = - V(y) \cdot {\mathcal W}_f, \qquad {\mathcal W}_f = \int_{y' \in {\mathcal Y}} V(y') \,  f(y')\, dy'  
\label{eq:kernel_3_hom} 
\end{eqnarray}
By (\ref{eq:mfl_equilibrium}) the Nash equilibrium is now depending on a parameter $W \in {\mathcal X}$. It is denoted by $M_W$ and given by (\ref{eq:Von-Mises}) with $\Phi(y) = - V(y) \cdot W$, i.e. 
\begin{eqnarray}
&&\hspace{-1cm}
M_W(y) = \frac{1}{Z_W} \exp \big( \frac{1}{d} (V(y) \cdot W) \big) \, \, \mbox{ with } \, \,  Z_W = \int_{y \in {\mathcal Y}}  \exp \big( \frac{1}{d} (V(y) \cdot W )\big) \, dy. 
\label{eq:Von-Mises_2}
\end{eqnarray}
Now, eq. (\ref{eq:mfl_equilibrium}) which defines a Nash equilibrium is replaced by a 'compatibility condition' deduced from  (\ref{eq:kernel_3}) and which expresses that 
$$ W = {\mathcal W}_{M_W}, $$
or equivalently 
\begin{eqnarray}
&&\hspace{-1cm}
W \int_{y \in {\mathcal Y}}  \exp \big( \frac{1}{d} (V(y) \cdot W) \big) \, dy = \int_{y \in {\mathcal Y}}  \exp \big( \frac{1}{d} (V(y) \cdot W) \big) \, V(y)  \, dy. 
\label{eq:Von-Mises_3}
\end{eqnarray}
The associated game is a potential game. Indeed, we introduce the potential energy 
\begin{eqnarray}
&&\hspace{-1cm}
{\mathcal U}(f) = - \frac{1}{2} |{\mathcal W}_f|^2. 
\label{eq:potential_VM}
\end{eqnarray}
According to definition (\ref{eq:fct_der}), we have, for all test functions $\phi(y)$, 
\begin{eqnarray}
\int_{y \in {\mathcal Y}} \frac{\delta {\mathcal U}(f)}{\delta f}(y) \, \phi(y) \, dy \nonumber
&=& - {\mathcal W}_f \cdot \int_{y \in {\mathcal Y}} \frac{\delta {\mathcal W}_f}{\delta f} \, \phi(y) \, dy \nonumber \\
&=& - {\mathcal W}_f \cdot \int_{y \in {\mathcal Y}} V(y) \, \phi(y) \, dy \nonumber \\
&=& \int_{y \in {\mathcal Y}} \Phi_f(y) \, \phi(y) \, dy, 
\label{eq:GatderivWf}
\end{eqnarray}
where, to pass from the first to the second line, we have used that, in the homogeneous configuration case, ${\mathcal W}_f$ is linear with respect to $f$. So, we get:
\begin{equation} 
\frac{\delta {\mathcal U}(f)}{\delta f}(y) = \Phi_f(y), 
\label{eq:pot_VM}
\end{equation}
which shows that the game with cost function $\mu_f(y)$ given by (\ref{eq:chempot_def}) is a potentiel game associated to potential ${\mathcal U}$. Its variational structure is associated to the following free energy functional (thanks to  (\ref{eq:entropy_def})  (\ref{eq:freeener_def}) and (\ref{eq:pot_VM})): 
\begin{eqnarray}
{\mathcal F}(f) = {\mathcal S}(f) - \frac{1}{2}|{\mathcal W}_f|^2.
\label{eq:freeener_VM}
\end{eqnarray}

In economics, one may be interested in the 'social cost' defined by
\begin{eqnarray}
{\mathcal C}(f) = \int_{y \in {\mathcal Y}} \mu_f(y) \, f(y) \, dy = {\mathcal S}(f) - \int_{y \in {\mathcal Y}} \Phi_f(y) \, f(y) \, dy,
\label{eq:soccost}
\end{eqnarray}
where the second equality comes from (\ref{eq:chempot_def}), (\ref{eq:entropy_def}). Now, taking $\phi = f$ in (\ref{eq:GatderivWf}) we have, for all $f \in {\mathcal P}_{\mbox{\scriptsize{ac}}}({\mathcal Y})$: 
\begin{eqnarray*}
\int_{y \in {\mathcal Y}} \Phi_f(y) \, f(y) \, dy = \int_{y \in {\mathcal Y}} \frac{\delta {\mathcal U}(f)}{\delta f}(y) \, f(y) \, dy . 
\end{eqnarray*}
But, noting that ${\mathcal U}(f)$ is a quadratic function of $f$, we have the following identity, which is valid for all degree $2$ homogeneous functions: 
\begin{eqnarray*}
\int_{y \in {\mathcal Y}} \frac{\delta {\mathcal U}(f)}{\delta f}(y) \, f(y) \, dy = 2  {\mathcal U}(f). 
\end{eqnarray*}
It follows that, for all $f \in {\mathcal P}_{\mbox{\scriptsize{ac}}}({\mathcal Y})$: 
\begin{eqnarray}
\int_{y \in {\mathcal Y}} \Phi_f(y) \, f(y) \, dy = -|{\mathcal W}_f|^2. 
\label{eq:order_parameter}
\end{eqnarray}
Then, the social cost (\ref{eq:soccost}) has the expression
\begin{eqnarray}
{\mathcal C}(f) = {\mathcal S}(f) - |{\mathcal W}_f|^2. 
\label{eq:soccost_2}
\end{eqnarray}
Note, that the free energy (\ref{eq:freeener_VM}) and social cost (\ref{eq:soccost_2}) differ by a factor $1/2$ in front of $|{\mathcal W}_f|^2$. This means that they differ except if ${\mathcal W}_f = 0$ (which is a kind of a measure of the social disorder). We note that a Nash equilibrium gives rise to a minimizer of the free energy instead of the social cost. The reason is that individual players make strategies without taking into account the cost of the freely available social infrastructure. To correct this discrepancy, one has to make players pay for the use of this infrastructure by e.g. assigning taxes. In the present cases, taxes would lead to a cost function equal to $\mu_f(y) - V(y) \cdot {\mathcal W}_f$. An example pertaining with city planning can be found in \cite{Blanchet_Mossay_Santambrogio}.

\subsection{Example: animal herding model}
\label{subsec:examples}

In this section, we consider a special case of the above one which describes the herding behavior a group of animals or a human crowd. The social space ${\mathcal X}$ coincides with the geographical space ${\mathbb R}^n$ (with $n=2$ (for crowds) or $n=3$ (for fish schools for instance)). The decision variable $y$ is the direction of the motion of the individuals and is such that  $y \in {\mathcal Y} = {\mathbb S}^{n-1}$, where ${\mathbb S}^{n-1}$ is the unit sphere of ${\mathbb R}^n$ endowed with the Lebesgue measure $dy$ (normalized such that the total measure of ${\mathcal Y}$ is equal to $1$). The function $V(y)$ relating the decision variable to the physical speed is independent of $x$ and simply given by: 
$$ V(y) = y, $$
where the speed (supposed uniform and independent of position) is normalized to $1$ through the non-dimensionalization procedure. We take the same kernel (\ref{eq:kernel_separate}) as before, which leads to the cost function 
\begin{eqnarray}
&& \hspace{-1cm} \Phi_f(x,y) = - y \cdot {\mathcal W}_f(x), 
\label{eq:kerex_1} 
\end{eqnarray}
with
\begin{eqnarray}
&& \hspace{-1cm} {\mathcal W}_f (x) = \int_{(x',y') \in {\mathbb R}^n \times {\mathbb S}^{n-1}} K(x,x') \, y' \, f(x',y') \, dx' \, dy' . 
\label{eq:kerex_2} 
\end{eqnarray}

In the homogenous configuration case (where we set $K=1$), the free energy is still given by (\ref{eq:freeener_VM}) where now ${\mathcal W}_f$ is given by 
\begin{eqnarray*}
&& \hspace{-1cm} {\mathcal W}_f = \int_{y' \in {\mathbb S}^{n-1}}  y' \, f(y') \, dy' .
\end{eqnarray*}
Then, from (\ref{eq:Von-Mises_2}), the Nash equilibrium is given by a so-called Von-Mises-Fischer (VMF) distribution
\begin{eqnarray}
&&\hspace{-1cm}
M_W(y) = \frac{1}{Z_W} \exp \big( \frac{1}{d} (y  \cdot W) \big), \qquad Z_W = \int_{y \in {\mathbb S}^{n-1}}  \exp \big( \frac{1}{d} (y  \cdot W) \big) \, dy ,  
\label{eq:Von-Mises_3.4}
\end{eqnarray}
 where $W \in {\mathbb R}^n$ is a solution of (\ref {eq:Von-Mises_3}). In the present context, this equation is written:
\begin{eqnarray}
&&\hspace{-1cm}
W \int_{y \in {\mathbb S}^{n-1}}  \exp \big( \frac{1}{d} (y  \cdot W) \big) \, dy = \int_{y \in {\mathbb S}^{n-1}}  \exp \big( \frac{1}{d} (y \cdot W) \big) \, y \, dy. 
\label{eq:Von-Mises_3.5}
\end{eqnarray}
By rotational symmetry and expressing the integrals in (\ref{eq:Von-Mises_3.5}) in polar coordinates, we can write 
\begin{eqnarray}
&&\hspace{-1cm}
W = |W| \,  \Omega, 
\label{eq:Omega_def}
\end{eqnarray}
where $\Omega \in {\mathbb S}^{n-1}$ is arbitrary. The quantity $\kappa = |W|/d$ satisfies:
\begin{eqnarray}
&&\hspace{-1cm}
c(\kappa) = d \, \kappa, 
\label{eq:Von-Mises_4}
\end{eqnarray}
with 
\begin{eqnarray}
&&\hspace{-1cm}
c(\kappa) = \frac{\int_0^\pi e^{\kappa \cos \theta} \, \cos \theta \, \sin^{n-2} \theta \, d \theta}{\int_0^\pi e^{\kappa \cos \theta} \, \sin^{n-2} \theta \, d \theta}, 
\label{eq:Von-Mises_5}
\end{eqnarray}
and $\cos \theta = y \cdot \Omega$. 
Then, the VMF distribution (\ref{eq:Von-Mises_3.4}) is more conveniently written $M_{\kappa \Omega}$ (where $\kappa \in {\mathbb R}_+$ is the concentration parameter and $\Omega$ is the mean direction) as: 
\begin{eqnarray}
&&\hspace{-1cm}
M_{\kappa \Omega}(y) = \frac{1}{Z_\kappa} \exp \big( \kappa (y  \cdot \Omega) \big), 
\label{eq:Von-Mises_6}
\end{eqnarray}
with $Z_\kappa$ given by the denominator of (\ref{eq:Von-Mises_5}). The quantity $c(\kappa)$ is the order parameter. It is an increasing function of $\kappa$ which satisfies $0 \leq c(\kappa) \leq 1$. When $c(\kappa) \approx 0$, then $M_{\kappa \Omega}(y)$ is nearly isotropic (i.e. $M_{\kappa \Omega}(y) \approx 1$). On the other-hand, when $c(\kappa) \to 1$, which happens when $\kappa \to \infty$, then $M_{\kappa \Omega}(y) \to \delta_\Omega(y)$ (see details in \cite{Degond_etal_JNLS12, Frouvelle_Liu_SIMA12}). 

Now, we look at the solutions of the compatibility condition (\ref{eq:Von-Mises_4}). This analysis has been performed in \cite{Degond_etal_JNLS12, Frouvelle_Liu_SIMA12}. We only summarize the final results in the following. 

\begin{theorem} \cite{Degond_etal_JNLS12, Frouvelle_Liu_SIMA12} 
(i) If $c'(0) = \frac{1}{n} \leq d$, then, the only solution of Eq. (\ref{eq:Von-Mises_4}) is $\kappa = 0$ and the associated equilibrium (\ref{eq:Von-Mises_6}) is the uniform distribution $M_0 = 1$. It is a stable equilibrium.  \\
(ii) If $c'(0) = \frac{1}{n} > d$, then, there exist exactly two solutions of Eq. (\ref{eq:Von-Mises_4}): $\kappa = 0$ and another solution denoted by $\kappa_d >0$. $\kappa_d$ is a strictly decreasing function of $d \in [0,\frac{1}{n}]$ onto $(+ \infty,0]$. The associated equilibria (\ref{eq:Von-Mises_6}) are the uniform distribution $M_0 = 1$ associated to $\kappa = 0$ and all VMF distributions $M_{\kappa_d \Omega}$ where $\Omega$ takes any value on the sphere ${\mathbb S}^{n-1}$. The uniform equilibrium is now unstable and the VMF equilibria $M_{\kappa_d \Omega}$ for all  $\Omega \in {\mathbb S}^{n-1}$ are the ground states of the free energy and are stable. 
\label{thm:phase_transition}
\end{theorem}

\noindent
We refer to \cite{Degond_etal_JNLS12, Frouvelle_Liu_SIMA12} for the precise mathematical statement of the stability result, as well as for rate estimates of convergence to the equilibria in the homogeneous configuration case. 

\begin{remark}
If $n=1$, then ${\mathbb S}^{n-1} = \{ -1, 1 \}$ and $c(\kappa) = \tanh (\kappa)$. The compatibility condition (\ref{eq:Von-Mises_4}) is the same as the mean-field equation in the Ising spin model for ferromagnetism \cite{Huang_StatMech87}.  
\label{rem:Ising}
\end{remark}

Now, we apply the coarse-graining procedure with Nash equilibrium closure developed in section \ref{sec:hydro} to this special case. To implement this closure, we first need to verify condition (\ref{eq:Phi_eps_exp}). For this purpose, we assume that 
\begin{eqnarray}
K(x,x') = \bar k \big( \big| \frac{|x-x'|}{\varepsilon} \big| \big), 
\label{eq:kernel_spheric}
\end{eqnarray}
with $\bar k(r)$: $r \in [0,\infty) \to {\mathbb R}_+$ a given kernel. The quantities associated to such kernel $K$ by  (\ref{eq:kerex_1}) and (\ref{eq:kerex_2}) are denoted by $\Phi_f^\varepsilon$ and ${\mathcal W}_f^\varepsilon$. For simplicity, we assume 
$$ \int_{x \in {\mathcal X}} \bar k  \big( \big| x \big| \big) \, dx = 1. $$
Then, inserting (\ref{eq:kernel_spheric}) into (\ref{eq:kerex_2}) and expanding in powers of $\varepsilon$, we get, for  $f \in {\mathcal P}_{\mbox{\scriptsize ac}}({\mathcal X} \times {\mathcal Y})$: 
\begin{eqnarray}
&& \hspace{-1cm} 
{\mathcal W}^\varepsilon_f (x) = \rho(x) \tilde {\mathcal W}_{\nu_x} + O(\varepsilon^2), 
\label{eq:kerex_3} 
\end{eqnarray}
with for all $\nu \in {\mathcal P}_{\mbox{\scriptsize ac}}({\mathcal Y})$, 
\begin{eqnarray}
&& \hspace{-1cm} 
\tilde {\mathcal W}_{\nu} = \int_{y' \in {\mathbb S}^{n-1}} y' \, \nu(y')  \, dy' . 
\label{eq:kerex_3.5} 
\end{eqnarray}
By inserting this expansion into (\ref{eq:kerex_1}), we get
\begin{eqnarray}
&& \hspace{-1cm}
\Phi_f^\varepsilon (x,y) = \Phi_{\rho, \, \nu_x} (y) + O(\varepsilon^2), 
\label{eq:kerex_4} 
\end{eqnarray}
where, for $\rho \in {\mathbb R}_+$ and $\nu \in {\mathcal P}_{\mbox{\scriptsize ac}}({\mathcal Y})$, we set
\begin{eqnarray}
&& \hspace{-1cm}
\Phi_{\rho, \, \nu} (y)  = - \rho \, y \cdot \tilde {\mathcal W}_\nu. 
\label{eq:kerex_5} 
\end{eqnarray}
We note that, in order to recover the homogeneous configuration setting of the beginning of this section, ${\mathcal W}_f$ must be replaced by $\rho \tilde {\mathcal W}_\nu$ in (\ref{eq:kerex_1}). It follows that the compatibility condition (\ref{eq:Von-Mises_4}) becomes
$$ c( \kappa ) = \kappa \, \frac{d}{\rho}, $$
and that the Nash equilibrium solutions are now the VMF distributions $M_{\kappa_{d/\rho} \Omega}$, with $\Omega \in {\mathbb S}^{n-1}$. The discussion of Theorem \ref{thm:phase_transition} is still valid provided that $d$ is replaced by $d/\rho$ everywhere. Therefore, there are two regimes corresponding to items (i) and (ii) in the statement of Theorem \ref{thm:phase_transition}. We successively describe the  models resulting from the application of Theorem \ref{thm:inhomogeneous} for these two regimes. 
\begin{itemize}
\item[(i)] Large noise or small density case: $\frac{d}{\rho} \geq \frac{1}{n}$. Then, the only Nash equilibrium being the isotropic distribution $M_0 = 1$, the macroscopic velocity $u$ as given by (\ref{eq:mean_vel_equi}), (\ref{eq:u}) is  
$$u = u[1] = \int_{y \in {\mathbb S}^{n-1}} y \, dy = 0, $$
by antisymmetry. The macroscopic equation (\ref{eq:mass}) reduces to 
$$\partial_t \rho  = 0 .$$
In order to get a meaningful macroscopic model, we must rescale time to diffusive scales. In this case, a diffusion approximation procedure leads to a nonlinear diffusion equation for $\rho$. Details can be found in \cite{Degond_etal_JNLS12}. 

\item[(ii)] Small noise or large density case: $\frac{d}{\rho} < \frac{1}{n}$. Then, we use the ground-state Nash equilibrium $M_{\kappa_{d/\rho} \Omega}$, where $\Omega \in {\mathbb S}^{n-1}$. In this case, $f$ is given by (\ref{eq:equi_nonhomo}), i.e.:
\begin{equation}
f(x,y,t) =  \rho(x,t) M_{\kappa_{d/\rho} \Omega(x,t)}(y) .
\label{eq:equi_nonhomo_VM}
\end{equation}
By the computations above, the mean velocity $u$ is given by
$$ u(x,t) = c(\kappa_{d/\rho}) \Omega(x,t) \not = 0.  $$
The Nash Equilibrium macroscopic closure equation (\ref{eq:mass}) gives
\begin{eqnarray}
&& \hspace{-1cm} \partial_t \rho + \nabla_x \cdot (c(\kappa_{d/\rho}) \rho \Omega) = 0 . \label{eq:mass_VM} 
\end{eqnarray}
We note that it does not provide an equation for $\Omega(x,t)$ yet and as such, would lead to an ill-posed problem. However, by using the concept of Generalized Collision Invariant (GCI) \cite{Degond_Motsch_M3AS08}, it is possible to derive the equation for $\Omega$. This equation reads:
\begin{eqnarray}
&& \hspace{-1cm} \partial_t \Omega + b(\rho) (\Omega \cdot \nabla_x) \Omega + \Theta(\rho) P_{\Omega^\bot} \nabla_x \rho = 0 , \label{eq:momentum_VM} 
\end{eqnarray}
where $b(\rho)$ and $\Theta(\rho)$ are real-valued functions of $\rho$ and $P_{\Omega^\bot} = \mbox{Id} - \Omega \otimes \Omega$ is the orthogonal projection of ${\mathcal X}$ onto the hyperplane space $(\mbox{Span} \{ \Omega \} )^\bot$ orthogonal to $\Omega$. The functions $b$ and $\Theta$ are not specified here. They are obtained through the application of the GCI to (\ref{eq:equi_nonhomo}). Details can be found in \cite{Degond_etal_JNLS12}. We note that because of the presence of the projection $P_{\Omega^\bot}$ the constraint $|\Omega|=1$ is propagated in time as soon as it is verified at time $t=0$. We note that the system is not in conservative form but, in some conditions, it can be shown to be well-posed \cite{Degond_etal_MAA13}.  
\end{itemize}

This example illustrates that the Nash Equilibrium closure can be effectively used to derive macroscopic closures. However, the mere mass conservation equation (\ref{eq:mass}) may be not be enough to provide a well-posed closed system and that additional techniques must be called for in order to find a closed system.

\setcounter{equation}{0}
\section{Conclusion and perspectives}
\label{sec:conclu}

In this paper we have provided a framework for the time evolution of a system of rational players in a non-cooperative anonymous game with a continuum of players (or Mean-Field Game) which collectively make their decision by choosing the steepest descent direction of the individual cost functions. Assuming that the individual actions are fast and localized in state space, we have derived a macroscopic dynamic which describes the large scale evolution of the parameters of the local Nash equilibria. In forthcoming works, we plan to apply this framework to various phenomena such as the evolution of the distribution of wealth in the economic neighborhood, opinion formation, social dynamics and collective decision making.


\end{document}